\documentclass[12pt]{article}
\usepackage{graphicx}
\usepackage{cancel}[]
\usepackage{amssymb,amsmath}
\usepackage{ifpdf}
\usepackage{hyperref}[]
\usepackage{fullpage}
\usepackage{color}
\usepackage{ulem}
\usepackage{verbatim}
\usepackage{float}
\ifpdf
    \DeclareGraphicsExtensions{.pdf}
\else
    \DeclareGraphicsExtensions{.eps}
\fi

\usepackage{graphicx,epsfig}
\usepackage{cancel}[]
\usepackage{amssymb,amsmath}
\usepackage{ifpdf}
\usepackage{hyperref}[]
\def\lsim{<\kern-2.5ex\lower0.85ex\hbox{$\sim$}\ }
\def\rsim{>\kern-2.5ex\lower0.85ex\hbox{$\sim$}\ }

\def\LAMBDABAR
{\hbox{$\lambda$\kern-0.52em\raise+0.45ex\hbox{--}\kern+0.2em}}

    \DeclareGraphicsExtensions{.eps}

\begin{document}

 \centerline{{\bf{The refractive index of the vacuum and the dark sector}}}
\vspace{0.2 in}
\centerline{A. C. Melissinos }
\centerline{\it{Department of Physics and Astronomy, University of Rochester }}
\centerline{\it{Rochester, NY 14627-0171, USA}}
\vspace{.10 in}

\centerline{June 7, 2017}

\vspace{.2 in}
In a recent publication \cite{PLB} it was reported that the analysis of preliminary data from the LIGO
interferometers shows no modulation at the sidereal frequency, improving the existing limits on the 
rotational invariance of the propagation of light by four orders of magnitude. The data however exhibit a strong modulation at twice the orbital frequency of the Earth's rotation around the sun. From an analysis of the data
in the context of the Standard Model Extension (SME) \cite{SME}, barring experimental error, it is found that the isotropic
trace coefficient $|\tilde{\kappa}_{tr}| = 3\times 10^{-9}$. This in turn implies that the
refractive index of light, while isotropic, differs from unity since $n = 1+\tilde{\kappa}_{tr}$.
This is a rather large value for the refractive index of the vacuum, and is in contradiction with
limits deduced from the observation of ultra high energy cosmic rays (UHECR) and of very high energy $\gamma$-rays (VHEGR), as first pointed out by Coleman and Glashow \cite{CG}, and discussed below.\\

While the refractive index is isotropic in a preferred frame, which we take to be the Sun-Centered-Celestial-Equatorial Frame (SCCEF) \cite{SME}, when 
boosted to the Earth frame the refractive index acquires a small anisotropy due to the change of the Earth's velocity as it orbits the Sun. Thus in the Earth frame the phase velocity differs in the arms of the interferometer. In addition, the tidal forces exerted on the Earth by the Sun and by the Moon lead to a differential gravity gradient along the arms, which also contributes to the observable phase shift of the light returning from the two arms.
This latter effect was clearly observed \cite{PLB} and is in agreement with the known frequencies 
of the tidal lines \cite{Melchior}. 

To transform $\tilde{\kappa}_{tr}$ to the Earth frame requires 
a double boost because $\tilde{\kappa}_{tr}$ is the trace of a $3 \times 3$ matrix (a 2-tensor); 
the boost depends on the orbital velocity of the Earth around the Sun\footnote{For this discussion
we ignore the
contribution of the rotational velocity of the Earth since it is an order of magnitude smaller than the  orbital velocity.}. As a result, the anisotropy that is observed in the Earth frame is of order $\Delta n 
\approx \beta_{\oplus}^2 \tilde{\kappa}_{tr}$. In fact, the observed difference in refractive
index between the two arms of the Hanford interferometer, after correcting for the tidal gradient,
 is of order $d n/n \approx 10^{-20}$. Because of the double boost, the resulting phase difference
 should be modulated at both the first and second harmonic of the orbital frequency. The data 
exhibit a very strong modulation at the second harmonic but no observable signal at
the first harmonic\footnote{This could be explained if the Lorentz violation is in the Fermion sector.},  suggesting that perhaps the observed effect is instrumental.\\

We now discuss the existing limits on $\tilde{\kappa}_{tr}$ or equivalently on the refractive index given in the Data Tables of 
Kostelecky and Russell \cite{Tables}. 
The strongest limits \cite{Klinkhamer, Schreck}, arise from the observation of ultra-high-energy cosmic rays and highly energetic $\gamma$-rays . When $|\tilde {\kappa}_{tr}|$$\neq$$0$ 
the phase velocity of light in vacuum, which we designate by $c_p$, differs from the ultimate 
velocity that can be attained by a material particle, and which we continue to designate\footnote{More precisely, $c$ provides the causal connection of space-time in the Minkowski
metric $(x^{\mu})= (x^0, \vec{x})=(ct,x^1,x^2,x^3)$.} by $c$ . The refractive index $n$ is related to
$\tilde{\kappa}_{tr}$ by 
\begin{equation} n=\frac{c}{c_p}=\sqrt{\frac{1 +\tilde {\kappa}_{tr}}{1-\tilde {\kappa}_{tr}}} \approx 1+\tilde {\kappa}_{tr}
 \end{equation}
where the approximation assumes $|\tilde{\kappa}_{tr}|\ll 1$; $\tilde{\kappa}_{tr}$ can take
either positive or negative values, and the observations bound both cases.\\

 Consider first the case $\tilde{\kappa}_{tr}>0 $ and thus $n>1$.

This would cause a charged particle to emit Cerenkov radiation, while traveling through 
vacuum, with consequent rapid energy loss until its energy is reduced below the threshold for
Cerenkov radiation. The condition for the onset of Cerenkov radiation \cite{Jackson}
is $\beta n \geqslant 1$, where $\beta = v/c$ is the normalized velocity of the particle,
say a proton, of mass $M_{p}$ and of energy $E=\gamma M_{p}$. It follows that the threshold
energy is obtained from
$$n^2\beta^2 = n^2(1-1/\gamma^2) =1,\qquad {\rm{or}}\qquad n^2/\gamma^2 = n^2-1$$
Since $\tilde{\kappa}_{tr} \ll 1$ we approximate $n^2-1 \approx 2\tilde{\kappa}_{tr}$
and $n^2 \approx 1$, to obtain for the threshold energy
\begin{equation} E_{threshold} = \frac{M_{p}}{\sqrt{2\tilde{\kappa}_{tr}}} \end{equation}
The appearance of UHECR with energy $E_{\mathrm{observed}}\approx 200\  \mathrm{EeV} = 2\times
10^{20}$ eV \cite{Auger}, and assuming for the primary cosmic ray, a mass $M=100$ GeV, sets a limit at the $2\sigma$ level \cite{Schreck} 
\begin{equation} \tilde{\kappa}_{tr} \leqslant 6\times 10^{-20} \end{equation}
\indent Next let $\tilde{\kappa}_{tr}<0$, and thus $n<1$.\ 

\indent In this case photons 
propagate with phase velocity $ c_p>c $. If their energy is $E_{\gamma}$, the three-momentum is 
$P_{\gamma} = E_{\gamma}/c_p = n E_{\gamma}/c <E_{\gamma}/c$, and therefore the photons are time-like, and can, and will decay into massive particles, as allowed by the conservation laws.
The lowest mass decay mode is $\gamma \rightarrow e^{+}e^{-}$ and the full momentum of 
the photon will be carried by the electron-positron pair. The kinematics imply a
threshold energy for this process\footnote{Note the analogy to Eq.2.} 
\begin{equation} 
 E_{\gamma} > \frac{2 m_e}{\sqrt{1-n^2}}  = \frac{2 m_e}{\sqrt{-2\tilde{\kappa}_{tr}}}
\end{equation}
and therefore, a lower bound on $\tilde{\kappa}_{tr}$
$$ \tilde{\kappa}_{tr} \geqslant - \frac{1}{2} \left(\frac{2 m_e}{E_{\gamma}}\right)^2 $$

\noindent Since a $56$ TeV $\gamma$-ray has been observed \cite{HEGRA}, this places a lower bound on 
$\tilde{\kappa}_{tr}$ at the $2 \sigma$ level \cite{Schreck}
\begin{equation} \tilde{\kappa}_{tr} \geqslant -2\times 10^{-16} \end{equation}
The limits (3) and (5) exclude by orders of magnitude the value of $\tilde{\kappa}_{tr}$ reported in \cite{PLB}, 
in the approximation that all the Fermion coefficients in the SME model are zero.\\

Finally we consider whether the presence of a dark photon sector can cause a vacuum refractive index for visible photons, for instance through kinetic mixing \cite{Holdom}.
The Lagrangian for such mixing is \begin{equation} \mathcal{L} = (1/2)(\vec{E}^2_{A} -
\vec{B}^2_{A}) +\mathcal{L_M} + (1/2)(\vec{E}^2_{D} - \vec{B}^2_{D}) \end{equation}
where \begin{equation} \mathcal{L_M} = (1/2)\zeta(F^A_{\mu \nu}F^{\mu \nu}_D) =
\vec{E}_A \cdot\vec{E}_D - \vec{B}_A \cdot\vec{B}_D \end{equation}
Here $A$ refers to the visible photons and $D$ to the dark sector photons, and $\zeta$
is a small dimensionless coefficient characterizing the coupling.
Instead, we propose  a mixing term of the form
\begin{equation} \mathcal{L_M} = (1/2)\xi(\vec{E}_A^2
\vec{E}_D^2 + \vec{B}_A^2\vec{B}_D^2)\end{equation} 
The proposed Lagrangian  of Eq.(8) is not renormalizable, manifestly
Lorentz violating and the parameter $\xi$ must have dimensions of $L^{-4}$.
However the Lagrangian\footnote{Terms higher than quadratic in the fields can be generated by loops and have 
been used by Weisskopf \cite{VFW} and by Heisenberg and Euler \cite{Heisenberg}
in their classic calculation of the birefringence of the vacuum.}   of Eq.(8) leads naturally to, see for instance \cite{Beresetskii},
\begin{equation}
\vec{D}_A = \frac{\partial \mathcal{L}}{\partial \vec {E}_A} = \vec{E}_A +
\xi \vec{E}_A E_D^2 = \vec{E}_A(1+\xi E_D^2) = \epsilon_A \vec{E}_A 
\end{equation}
\begin{equation}
\vec{H}_A = - \frac{\partial \mathcal{L}}{\partial \vec {B}_A} = \vec{B}_A -
\xi \vec{B}_A B_D^2 = \vec{B}_A(1-\xi B_D^2) = \mu_A^{-1} \vec{B}_A 
\end{equation}
We thus obtain for the dielectric constant $\epsilon_A$ and permeability $\mu_A$
of the visible field \begin{equation} \epsilon_A = 1+\xi E_D^2 \qquad \qquad 
\mu_A = 1+\xi B_D^2 \end{equation} 
and for the refractive index of the visible photons
\begin{equation} n =\sqrt{\epsilon_A \mu_A} = 1+ \frac{\xi}{2}(E_D^2 +B_D^2)
\end{equation}
In the reference frame the refractive index is isotropic and does not depend on the orientation of the dark or visible fields. It depends only on the EM energy density $\rho^{EM}_D = (E_D^2
+ B_D^2)$ of the dark photon field. 
The dark matter density in our galaxy is presumed to be $\rho_D \approx 300 \ \mathrm
{MeV/cm^3}$ and has a thermal velocity distribution \cite{Turner}. If we assume that 
the dark photon energy density $\rho_D^{EM}$ is in equilibrium with $\rho_D $ in the dark matter
rest frame, we can set $\rho_D^{EM}=\rho_D$. In the Earth frame $\rho_D^{EM}$ will have approximately the same value as in the reference frame because the relative velocities are small.\\

I thank Ashok Das, Alan Kostelecky and Matt Mews for many discussions and helpful advice.


\end{document}